\documentclass[twoside,fleqn]{article}
\usepackage{graphicx}
\usepackage{epsfig}
\usepackage{espcrc2}

\newcommand{\eq}{\begin{equation}}
\newcommand{\en}{\end{equation}}
\newcommand{\eqa}{\begin{eqnarray}}
\newcommand{\ena}{\end{eqnarray}}

\newcommand{\AmS}{{\protect\the\textfont2
  A\kern-.1667em\lower.5ex\hbox{M}\kern-.125emS}}

\title{A nonperturbative determination of $C_A$ and the scaling of $f_\pi$ and
$m^{\overline{MS}}$} \author{UKQCD Collaboration, presented by S.~Collins\address{Dept. of Physics and Astronomy,
    Glasgow University, Glasgow, G12 8QQ, Scotland}\thanks{In collaboration with C.~Davies~(Glasgow University), G.~P.~Lepage~(Cornell University),
and J.~Shigemitsu~(Ohio-State University)}}

\begin{document}

\begin{abstract}
We calculate the $O(a)$ improvement coefficient for the axial-vector
current using the nonperturbative method proposed by the LANL group.
Results for the coefficient in the range $\beta=5.93$ to 6.2 are
presented. We find $c_A$ is close to the 1-loop tadpole-improved
perturbative value. In addition, scaling of the pseudoscalar decay
constant and renormalised quark mass is improved compared to that
obtained using the (larger) $c_A$ values obtained by the ALPHA
collaboration.
\end{abstract}

\maketitle

\section{Introduction}
Symanzik improvement offers reduced discretisation errors for masses
and matrix elements at finite lattice spacing but requires the
calculation of the coefficients which accompany the improvement terms.
In the light quark regime these coefficients can be calculated
nonperturbatively by imposing the chiral Ward identities on the
lattice.

\begin{figure}
\begin{center}
\includegraphics[width=5.0truecm,angle=-90]{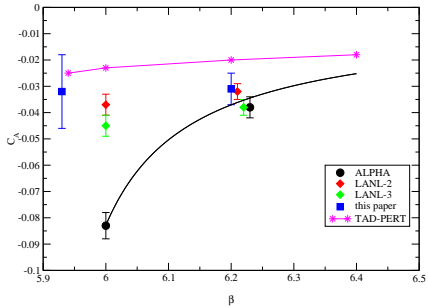}
\vspace{-1cm}
\caption{Summary of $c_A$ determinations. LANL-2 and LANL-3 refer
to $c_A$ determined using 2-point and 3-point lattice derivatives
respectively, as described in reference~\protect\cite{lanl}.}
\label{ca}
\end{center}
\vskip -0.75truecm
\end{figure}

$c_A$ is the $O(a)$ improvement coefficient for the axial-vector
current.
\begin{equation}
A_4 \rightarrow A_4^I = A_4 + ac_A\partial_4 + O(a^2)
\label{imp}
\end{equation}
It has been calculated nonperturbatively by the ALPHA
collaboration~\cite{alpha} and LANL group~\cite{lanl}. The results are
summarized in figure~\ref{ca}. There is a significant discrepancy in
the values of $c_A$ at smaller $\beta$. This might be accounted for by
the $O(a)$ ambiguity that exists in $c_A$ when it is determined
nonperturbatively.  Nevertheless, larger values of $c_A$ lead to
stronger lattice spacing dependence of $f_\pi$ and the quark mass, and
increase the difficulty of performing a continuum extrapolation. In
addition, the ALPHA results have a very different $\beta$ dependence
compared to the 1-loop perturbative values, which suggests they
contain a large nonperturbative contribution. Our aim was to
investigate how well $c_A$ is determined using the LANL method, which
only requires conventional analysis, rather than Schr\"odinger
functional techniques of the ALPHA collaboration.

$c_A$ is determined nonperturbatively using the PCAC relation. In
euclidean space, for zero momentum
\begin{equation}
<\partial_4A_4(x)J> = 2m_{PCAC}<P(x)J>\label{pcac}
\end{equation}
where $m_{PCAC}$ is the PCAC quark mass, $P=\bar{\psi}\gamma_5\psi$
and $J$ is any operator with the pseudoscalar quantum numbers.
Defining $r_J(t)=\frac{<\partial_4 A_4(t) J>}{<P(t)J>}$,
equation~\ref{pcac} becomes on the lattice
\begin{eqnarray}
 r_J(t) = m(t) &=& 2m_{PCAC}+O(a).
\end{eqnarray}
The size of the $O(a)$ terms depends on the number of states
contributing to $r_J(t)$. At early times~($t_{ex}$), the significant
contribution from excited states means the $O(a)$ term is larger than
at later times when the ground state dominates~($t_{gs}$). The
improvement of equation~\ref{imp} removes the $O(a)$ terms for all
states and hence the two masses
\begin{eqnarray}
r_J(t_{gs}) + ac_As_J(t_{gs}) & = & m_{imp}(t_{gs}) = \nonumber \\ & & 2m_{PCAC} + O(a^2).\\
r_J(t_{ex}) + ac_As_J(t_{ex}) & = & m_{imp}(t_{ex}) = \nonumber \\ & & 2m_{PCAC} + O'(a^2).
\end{eqnarray}
where $s_J(t) = \frac{<\partial_4^2 P(t) J>}{<P(t)J>}$, are equal
up to higher order terms. We can use these equations to solve for $c_A$:
\begin{eqnarray}
 -\frac{1}{a} \frac{r(t_{ex}) - r(t_{gs})}{s(t_{ex})-s(t_{gs})} & \equiv & c_A \label{ratca}
\end{eqnarray}
The LANL method involves performing a fit to
$r_J(t)+ac_As_J(t)=constant=2m$ over the fitting range $t_{ex}$ to
$t_{gs}$ and is numerically equivalent to equation~\ref{ratca}; $c_A$
and $2m$ are extracted from the fit. The advantage of performing a fit
is that one can test the ansatz with the $\chi^2$.

The $O(a)$ ambiguity in $c_A$ is due to $O(a^2)$ terms in the
axial-vector current, the pseudoscalar current and the light quark
action. We define $c_A$ in the chiral limit, but at finite quark mass
there is an additional source of $O(a)$ error if standard, symmetric,
$O(a^2)$ lattice temporal derivatives are used. This error can be
removed by using improved $O(a^4)$ lattice derivatives. While, both
choices of lattice derivative lead to consistent values for $c_A$ in
the chiral limit, we found it advantageous to use improved derivatives
as this weakened the dependence of $c_A$ on the quark mass and made
the chiral extrapolation easier. We also found a wider window in
$t_{ex}$ from which $c_A$ can be extracted.

\section{Results for $c_A$}
We calculated $c_A$ on the UKQCD Collaboration quenched data set, at
$\beta=5.93$, $6.0$ and $6.2$; simulation details can be found
in~\cite{ukqcd}. The best analysis was possible at
$\beta=5.93$.

\begin{table}
\begin{center}
\begin{tabular}{cc|c}\hline
$t_{ex}$ & $\kappa=0.1327$ & $\kappa=\kappa_c$ \\\hline
\multicolumn{3}{c}{$FL$}\\\hline
4  & -0.037(2)& -0.050(3) \\
5  & -0.035(4) & -0.052(8) \\
6  & -0.029(9) & -0.022(19) \\
7  & -0.025(15) & - \\
8 &  -0.066(264) & -\\\hline
\multicolumn{3}{c}{$LL$}\\\hline
6  & -0.024(6) & -0.032(14) \\
7  & -0.022(11) & - \\
8  & -0.046(30) & - \\\hline
\end{tabular}
\caption{$c_A$ values extracted for $\beta=5.93$ at finite
quark mass and extrapolated to the chiral limit. For all results shown
$Q$ for the fit is $>0.01$. At $\kappa_c$ and higher $t_{ex}$ the
statistical errors dominate and the results are not
shown. $t_{gs}=14$.}
\label{tab}
\end{center}
\vskip -0.75truecm
\end{table}

Table~\ref{tab} presents the results for $c_A$ at $\beta=5.93$ for
``fuzzed'' source~\cite{ukqcd} and $LL$ correlators, where improved
lattice derivatives have been used. We see that there is agreement
between the results from different smearings and the results are
stable with $t_{ex}$. However, there is only a small window of
timeslices for which $Q$ for the fit is reasonable and before noise
dominates. No significant change in $c_A$ was found when $t_{gs}$
was varied.

We take $c_A=-0.032(14)$ as our final value for this $\beta$. In
choosing a value with a conservative error we aim to take into account
the residual $O(a)$ ambiguity in $c_A$ and some of the other
systematic errors.  In principle all results in the table are valid
estimates of $c_A$. However, choosing a result with a much smaller
error would require much more care in keeping the improvement
conditions applied fixed when $c_A$ is recalculated at other $\beta$s;
thus ensuring, for example, any residual $O(a)$ error in $c_A$
vanishes when the continuum limit is taken. In particular, with the
LANL method one would need to keep the relative contribution of ground
state to excited states at $t_{ex}$~(in physical units) constant and
this requires tuning of the smearing function, which was
not possible in this study.

Repeating the analysis at $\beta=6.0$ we found evidence of significant
finite volume dependence. $c_A$ decreased from $-0.039(6)$ at
$\kappa=0.13344$ on a $16^3$ lattice to $-0.003(10)$ on a $32^3$
lattice. The $16^3$ lattice is $16\%$ smaller in extent than that used
at $\beta=5.93$. Due to problems with chiral extrapolations, our final
results as a function of $t_{ex}$ either had too small a statistical
error to reflect the change in volume, or had too large an error to be
useful. Hence, we disregard the results at this $\beta$ in the later
analysis. At $\beta=6.2$, the lattice size is only $6\%$ smaller, and
we obtain $c_A=-0.031(5)$. We assume the latter error is sufficient to
compensate for the small change in volume from $\beta=6.2$ to $5.93$.

\section{Scaling of $f_\pi$ and $m^{\overline{MS}}$}
\begin{figure}
\begin{center}
\epsfxsize=6.0truecm\epsffile{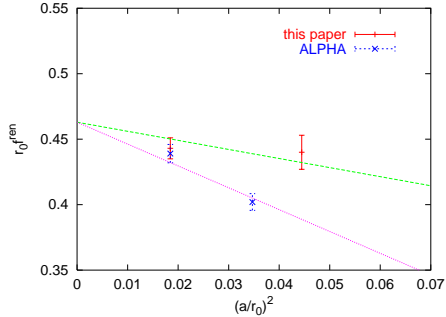}
\vspace{-1cm}
\caption{Scaling of the renormalised $PS$ decay constant in units of $r_0$.}
\label{fscal}
\end{center}
\vskip -0.75truecm
\end{figure}

We calculate the renormalised pseudoscalar decay constant, given by
\begin{eqnarray}
f^{ren}  =  Z_A(1+ab_Am_q)(f^{(0)}+ac_Af^{(1)})& &\\
f^{(0)}=<0|A_4|P>,  f^{(1)}=<0|\partial_4P|P> & &
\end{eqnarray}
using our results for $c_A$ and nonperturbative determinations of
$Z_A$ and the perturbative results for $b_A$; $f^{ren}$ is extracted
at a reference mass $(r_0M_{PS})^2=3.0$. This is compared in
figure~\ref{fscal} with the results using the ALPHA collaboration
values for $c_A$. We see that scaling is much better when our smaller
values of $c_A$ are used. The simultaneous fit to all 4 points only
has a marginal $\chi^2=2.4$. More points and higher orders in $a$
would be needed to include the ALPHA $c_A$ values for $f^{ren}$.

\begin{figure}
\begin{center}
\epsfxsize=6.0truecm\epsffile{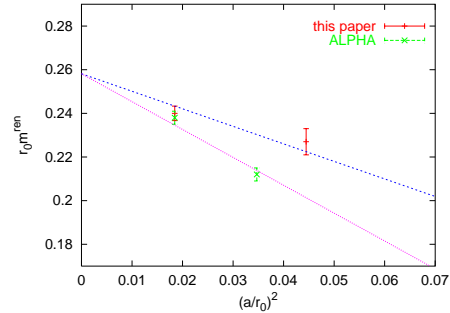}
\vspace{-1cm}
\caption{Scaling of the renormalised quark mass in units of $r_0$.}
\label{mscal}
\end{center}
\vskip -0.75truecm
\end{figure}

The renormalised quark mass, at the same reference mass, was
calculated using the method of reference~\cite{squark}. The
renormalisation-group invariant mass is computed from the PCAC mass using
the nonperturbatively determined renormalisation factor $Z_M$:
$\hat{M}=Z_Mm_{PCAC}^{imp}$ The $\overline{MS}$ mass at $2$~GeV is
then obtained using 4-loop perturbation theory. Figure~\ref{mscal}
shows that the scaling is better for smaller values of $c_A$; the
$\chi^2$ for the combined fit is $3.3$.

\section{Conclusions}
We implemented the LANL method for determining $c_A$ in the range
$\beta=5.93$ to $6.2$. Figure~\ref{ca} shows our results are
consistent with those of the LANL group. In particular, $c_A$ does not
change rapidly with $\beta$ and is close to the 1-loop
tadpole-improved value~(using $\alpha_P(1/a)$). At $\beta=6.0$,
evidence of a finite volume dependence in $c_A$ indicated care must be
taken in keeping the physical volume in the simulation fixed when
changing $\beta$ with this method.

The scaling of both $f_\pi$ and $m^{\overline{MS}}$ are improved using
our, smaller, values of $c_A$ compared to the values obtained by the
ALPHA collaboration.

S.~Collins acknowledges the support of a Royal Society of Edinburgh
fellowship. 

\end{document}